\documentclass[preprint]{revtex4}
\newcommand{\be}{\begin{equation}}
\newcommand{\ee}{\end{equation}}
\newcommand{\bea}{\begin{eqnarray}}
\newcommand{\eea}{\end{eqnarray}}
\usepackage{graphicx,color}
\usepackage{epsfig}
\usepackage{bm}
\usepackage{float}
\usepackage{amssymb}

\begin{document}
\title{Free Energy Evaluation in Polymer Translocation via Jarzynski Equality}
\author{Felipe Mondaini\footnote{Email address of corresponding author: fmondaini@if.ufrj.br}}
\affiliation{Centro Federal de Educa\c{c}\~ao Tecnol\'ogica Celso Suckow da Fonseca, \\
Petr\'opolis, 25.620-003, RJ, Brazil}
\author{L. Moriconi\footnote{Email address: moriconi@if.ufrj.br}}
\affiliation{Instituto de F\'isica, Universidade Federal do Rio de Janeiro, \\
C.P. 68528, 21945-970, Rio de Janeiro, RJ, Brazil}
\begin{abstract}
We perform, with the help of cloud computing resources, extensive Langevin simulations which provide free energy estimates for unbiased three dimensional polymer translocation. We employ the Jarzynski Equality in its rigorous setting, to compute the variation of the free energy in single monomer translocation events. In our three-dimensional Langevin simulations, the excluded-volume and van der Waals
interactions between beads (monomers and membrane atoms) are modeled through a repulsive Lennard-Jones (LJ) potential and consecutive monomers are subject to the Finite-Extension Nonlinear
Elastic (FENE) potential. Analysing data for polymers with different lengths, the free energy profile is noted to have interesting finite size scaling properties.
\end{abstract}
% \pacs{87.15.Vv, 05.40.-a, 87.15.A-}
%\pacs{}
\maketitle
\section{Introduction}
The process of polymer translocation occurs in many biological and biotechnological phenomena. It has received great attention in both experimental and theoretical studies in recent years due to its important role in many crucial biological processes, such as mRNA translocation across a nuclear pore complex~\cite{suntharalingam2003}, drug delivery~\cite{Tseng}, injection of DNA from a virus head into a host cell and gene therapy~\cite{Szabo, Hanss}. However due to the complexity of the interactions involved, especially between the pore and the membrane, computer simulations have been widely used as a fundamental research tool. Most of the numerical studies can be classified into the topical issues of (i) translocation driven by chemical potential gradients~\cite{Kasianowicz, Grosberg, Huopaniemi, Luo2007, Wei2007, Luo2008}, (ii) translocation driven by external forces~\cite{Bhattacharya2009}, and (iii) unbiased translocation~\cite{Gauthier2009, Panja2007}.

The first observation of polymer translocation was done experimentally by Kasianowicz \emph{et al.}~\cite{Kasianowicz}, using ssDNA fragments driven through a narrow $\alpha$-hemolysin nanopore in a biomembrane by application of an external voltage. Kasianowicz \emph{et al.}~\cite{Kasianowicz} observed that the mean translocation time, $\langle \tau \rangle$, scales linearly with polymer length, $N$, and inversely with the potential difference across the pore. Putting together the Fokker-Planck approach and classical nucleation theory,
Sung and Park~\cite{Sung} and Muthukumar~\cite{Muthukumar199910371} theoretically predicted that the translocation time scales with polymer chain length $N$ as $\langle \tau \rangle \sim N$ for driven translocation and $\langle \tau \rangle \sim N^2$ for the case of nondriven translocation.
In 2002, Chuang \emph{et al}~\cite{Chuang2002} pointed out that the estimate of the translocation time, $\tau_{trans}\sim N^2$, based on the Fokker-Planck equation for the first-passage time over the membrane entropic barrier couldn't be correct since $\tau_{trans}$ cannot be smaller than the Rouse time, which scales as $\tau_{R}\sim N^{1+2\nu}$. This prediction was confirmed for many simulations in 2D and in 3D and had a special attention in Ref.~\cite{Mondaini2}, where polymers with a relatively large number of monomers have been studied.

From the statistical physics point of view, the polymer translocation problem can be seen as a kind of tunneling process over an entropic barrier~\cite{Muthukumar199910371}. The entropic barrier results from the decrease in the number of allowed conformations which takes place in the presence of the membrane. For a polymer of length $N$, with one end fixed on a wall, the total number of conformations, $Z$, can be written as $Z(N)=\overline{z}^NN^{\gamma-1}$, where $\overline{z}$ is essentially the effective coordination number for the orientation of the adjacent bonds, and $\gamma$ is a scaling exponent. This exponent depends on the nature of the polymer and the background fluid, and also on boundary conditions related to the specific membrane geometry. The parameter $\overline{z}$ can be written as $\exp(-\mu/k_{B}T)$, where $\mu$ is the chemical potential. The Helmholtz free energy $F_{N}$ of the chain is related to the partition function $Z(N)$ as $F_{N}=-k_{B}T \ln Z(N)$. We have, then,
\be
\frac{F_{N}}{k_{B}T}=\frac{\mu N}{k_{B}T}-(\gamma-1)\ln(N) \ . \
\label{muthu1}
\ee

In the translocation problem, the polymer chain undergoes translocation through the hole from region I (cis) to region II (trans). In a intermediate state during the translocation, let there be $m$ segments in the region II and $N-m$ segments in the region I. The total partition function will be the product of the partition functions for the two tails, $Z=Z_{I}(N-m)Z_{II}(m)$, and in the same way as before, the free energy of the chain $F(m)$, with $m$ monomers in the region II can be written as:
\be
\frac{F(m)}{k_{B}T}=(1-\gamma_{2}^{'})\ln(m) +(1-\gamma_{1}^{'})\ln(N-m)-m\frac{\Delta \mu}{k_{B}T} \ , \
\label{muthu2}
\ee
where $\gamma_{1}^{'}$ and $\gamma_{2}^{'}$ are the values of $\gamma^{'}$ in the regions I and II, respectively, and $\Delta \mu\equiv \mu_{1}-\mu_{2}$ is the chemical potential difference across the membrane. In our study, we focus on a homopolymer surrounded by some homogeneous solution. We take, thus, $\gamma_{1}^{'}=\gamma_{2}^{'}=\gamma=0.5$ and $\Delta \mu=0$.

It is important to observe that this approach employs the quasi-equilibrium approximation once it is assumed that unbiased translocation is sufficiently slow, so that the sections of the polymer outside the nanopore remain in conformational equilibrium. The validity of the quasi-equilibrium approximation for polymer translocation has been challenged in both nondriven~\cite{Panja2007, Dubbeldam2007} and driven~\cite{sakaue2007, sakaue2010, rowghanian2011} translocation. Some authors have suggested that polymer translocation is a strong non-equilibrium process and having this in mind they have proposed alternative descriptions~\cite{lehtola2009, lehtola2008, polson2013}. Unfortunately, these methods are essentially based on Monte Carlo and molecular dynamics for short polymer chains. Since it is a difficult task to derive analytical expressions for free energy in the translocation problem, the use of numerical simulations within a cloud computing framework, together with modern non-equilibrium theorems comes as a promising way
to pursue.

In this article we provide numerical evaluations of the free energy in the polymer translocation problem via the Jarzynski Equality (JE)~\cite{Jarzynski, Latinwo, Seifert}. With the help of cloud computing resources~\cite{teste} we have been able to introduce statistical ensembles and study polymer chains which are considerably larger than the ones usually taken in the literature. In Sec. II, we briefly review the JE and its applications. In Sec. III, we discuss the numerical details related to the Langevin simulations. In Sec. IV, we apply the JE to the problem of polymer translocation, where we then obtain free energy profiles. Finally in Sec. V, we summarize our results and point out directions of further research.

\section{Jarzynski Equality}
C. Jarzynski derived, in a straightforward and seminal paper~\cite{Jarzynski}, a relation between the path dependent work and the path independent free energy change of a thermodynamical system, as given from the following sum rule,
 \be
 \int_{-\infty}^{\infty} dW \rho(W)e^{-\beta W}=\langle e^{-\beta W} \rangle=e^{-\beta \Delta F} \ , \
 \ee
where $\rho(W)$ denotes the probability density function of the work $W$ that is perfomed on the system, and $\Delta F=F_{f}-F_{i}$, denotes the difference between the free energies $F_{i}$ and $F_{f}$ of the initial thermal equilibrium state and the final thermal equilibrium state of the system, respectively. It is important to emphasize that the non-equilibrium processes involved in the definition of the JE all start at some well-defined equilibrium thermodynamic state with temperature $T=(k_{B}\beta)^{-1}$. The processes are guided by the action of forces related to the time-dependent parameters that characterize the Hamiltonian of the system.

The JE provides a way to compute the equilibrium quantity $\Delta F$ from an ensemble of finite-time, nonequilibrium measurements of the work done on the system, as given by the statistical average $\langle e^{-\beta W} \rangle$. Any application of the JE requires repeating experiments (real or numerical) with the same protocol a great number of times in order to generate sufficient statistics. Some care is necessary when working with the JE since the work average can have a very slow convergence, in particular if we deal with macroscopic states.

The validity of this equality, having in mind determinations of the free energy change $\Delta F$ was sucessfully demonstrated in various experimental systems such as, for instance, classical oscillators~\cite{Bena2005, Blicke2006} and single molecules~\cite{Liphardt2002}. In a particular class of single-molecule experiments suggested by Hummer and Szabo~\cite{Hummer2001}, where the end-to-end distance of the molecule is controlled by means of optical traps, the molecule is attached to a glass slab on one end and a polystyrene bead on the other, which in turn interacts with the external laser trap through an effectively harmonic potential. In this case, the external control parameter in the JE specifies the position
of the confining potential. The central idea of this paper is to use the JE approach in the polymer translocation problem, as it will be discussed in detail in Sec. IV.
\section{Computational Strategy}
In our three-dimensional Langevin simulations, the excluded-volume and van der Waals interactions between beads (monomers and membrane atoms) separated by a distance $r$ are modeled through a repulsive Lennard-Jones (LJ) potential~\cite{lennard1937}, $U_{LJ}(r)$ with cutoff at length $2^{1/6}\sigma$, where $\sigma$ is the bead diameter:
\be
U_{LJ}(r) = \left\{ \begin{array}{ll}
4 \epsilon [(\sigma/r)^{12}-(\sigma/r)^6]+\epsilon  \ , \
{\hbox{ if }} r \leq 2^{1/6} \sigma \ , \ \\
0  \ , \  {\hbox{ if }} r > 2^{1/6} \sigma \ . \
\end{array}\right.
\ee
Besides the LJ potential, consecutive monomers are subject to the
Finite-Extension Nonlinear Elastic (FENE) potential,
\be
U_{F}(r) = - \frac{1}{2} k R_0^2 \ln [ 1 -(r/R_0)^2] \ . \
\ee
From the above definition, it is clear that the FENE potential does not allow the distance between consecutive monomers to become larger than $R_0$.

\begin{figure}[tbph]
\centering
\includegraphics[scale=0.6]{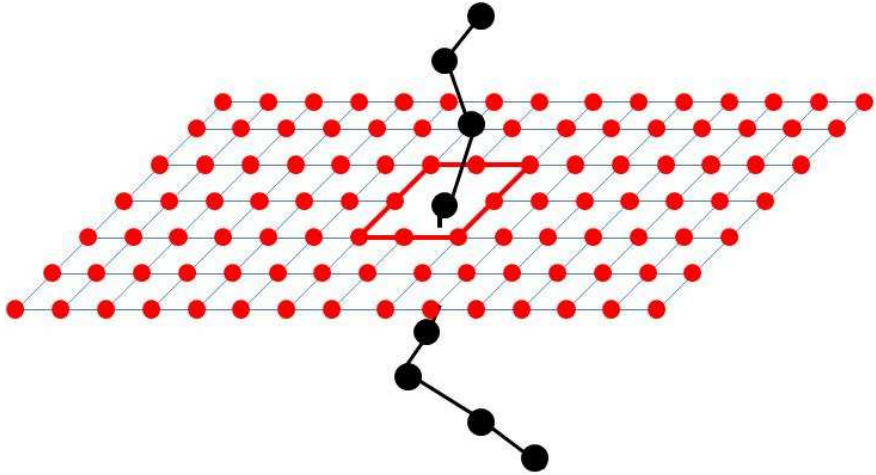}
\caption{A polymer translocates through a membrane pore in three-dimensional space. The pore is created by the remotion of a single atom at the center of an 80 x 80 monoatomic square lattice membrane.}
\label{pol2}
\end{figure}

We have studied polymers with sizes up to 100 monomers, which translocate through a pore created by the remotion of a single atom at the center of an 80 x 80 monoatomic square lattice membrane similar to the one shown in Fig.~\ref{pol2}.

The translocation process can be dynamically described by the following Langevin equations,
\be
m \frac{d^2 \vec r_i}{dt^2} = - \sum_{ j \neq i} \vec \nabla_{r_i}  [U_{LJ}(r_{ij}) + U_F(r_{ij})] -\xi \frac{d \vec r_i }{dt} + \vec F_i(t) \ , \ \label{eq-motion}
\ee
where $r_{ij} = |\vec r_i - \vec r_j|$, $\xi$ is the dissipation constant and $\vec F_i(t)$ is a gaussian stochastic force which acts on the monomer with label $i$, being completely defined from the expectation values
\bea
&&\langle \vec F_i(t) \rangle = 0 \ , \ \nonumber \\
&&\langle [\hat n \cdot \vec F_i(t)] [\hat n' \cdot \vec F_j(t') ] \rangle = 2 \hat n \cdot \hat n' k_B T \xi \delta_{ij} \delta(t-t') \ . \
\eea
Above, $\hat n$ and $\hat n'$ are arbitrary unit vectors, and $k_B$ and $T$ are the Boltzmann constant and the temperature, respectively. By means of a suitable regularization of the stochastic force, we have implemented a fourth-order Runge-Kutta scheme for the numerical simulation of the Langevin Equations (\ref{eq-motion}). Our simulation parameters are: $\epsilon=1.0$, $\sigma=1.0$ ($\sigma$ is also identified with the membrane lattice parameter), $\xi =0.7$, $k=7 \epsilon/\sigma^2$, $R_0 = 2 \sigma$, $k_B T = 1.2 \epsilon$. The simulation time step is taken as $3 \times 10^{-3}t_{LJ}$, where $t_{LJ} \equiv \sqrt{m \sigma^2 / \epsilon}$ is the usual Lenard-Jones time scale. In the most general case, the initial configuration of the polymer has $m$ monomers on the trans-side of the membrane and $N-m$ on the cis-side. Translocation is allowed to start only after an initial stage of thermal equilibrium is reached for the whole polymer.

\section{Free Energy Evaluations}
In order to get an operational understanding of the JE in the polymer translocation context, let us consider a situation where the monomer position at the pore is determined by an arbitrary potential that acts like an optical trap. This potential, for our purposes, is a time-dependent modulated well, $U(r,t)$, which induces the translocation of a single monomer,
\be
U(r,t)=Ae^{-br^2}(\sin[cz- \lambda(t)])^{2} \ . \ \label{Upot}
\ee
Above, $A$, $b$ and $c$ are positive constants, and $\lambda(t) \equiv \pi v t $ is the control parameter in our application of the JE, where $v>0$ is the speed of the minimum potential location along the $\hat{z}$ direction (perpendicular to the membrane plane). As it is described in more detail below, a single monomer translocation event takes place when $\lambda(t)$ is changed from $\lambda =0$ to $\lambda = \pi$.

A relevant issue in applications of the JE is to determine how invasive is the protocol procedure put forward in the evaluation of free energy variations. It is possible, in some specific cases related to the use of harmonic traps, to generalize the JE so that one can eliminate such sources
of error~\cite{Hummer2001}. In our particular analysis, we have designed the trapping potential at the pore, Eq. (\ref{Upot}), to be minimally invasive. In fact, the energy fluctuations of the monomer confined at the pore by the potential (\ref{Upot}) are of the order of $k_BT$ for a polymer chain of length $N$. Once the spatial range of the trapping potential is smaller than the mean distance between consecutive monomers, it is reasonable to assume that it will not lead to important entropic effects. The error in the free energy variations (measured in $k_BT$ units) associated to the specific protocol given in Eq. (\ref{Upot}) is, therefore, of the order of unity. We have checked, {\it{a posteriori}}, that this gives a neglible contribution (smaller than $1 \%$) to the computed free energy variations.

\begin{figure}[]
\centering
\includegraphics[scale=0.6]{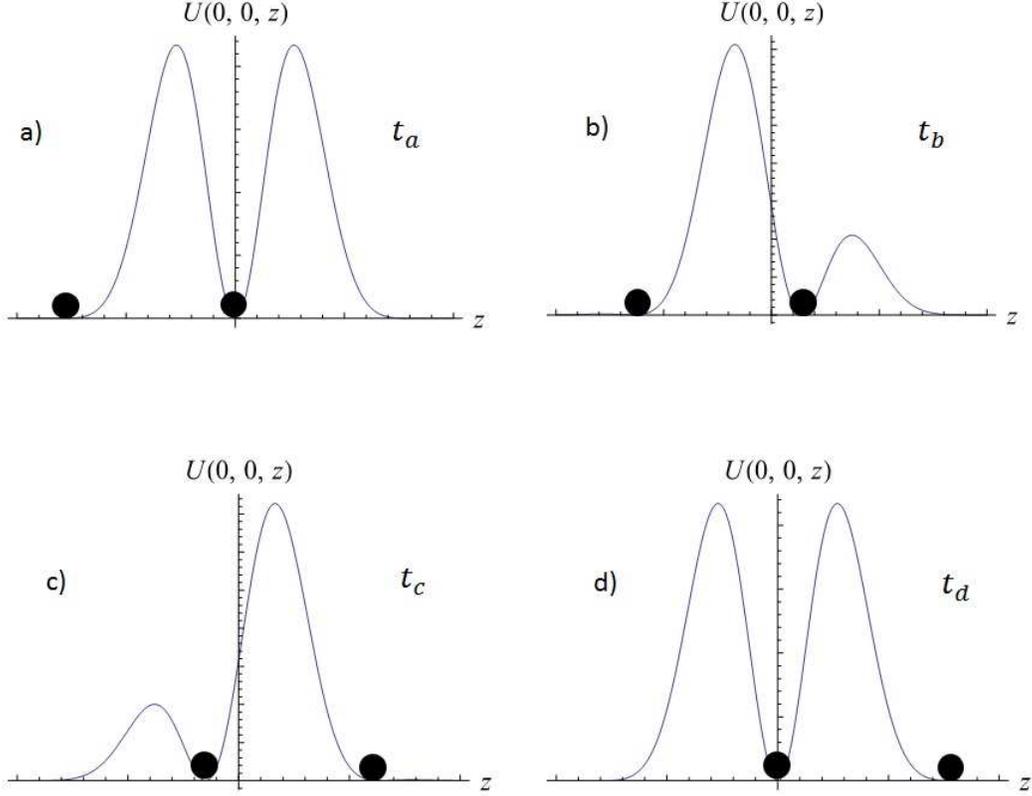}
\caption{(a) At $t_a$, soon after thermalization, the polymer has a monomer attached to the pore, preventing other monomers from translocating. (b) At $t_b$, the monomer which was previously attached to the pore moves along with the potential well, which in its turn develops a smaller right peak. (c) At $t_c$, the right monomer is free, while the left one is imprisoned by the potential. (d) At $t_d$, the left monomer is finally brought the pore, ending the single-monomer translocation induced by the action of the time-dependent potential.}
\label{potencial}
\end{figure}

We show, in Fig.~\ref{potencial}, snapshots of the time dependent potential (\ref{Upot}) at four time instants. A monomer is initially placed at the pore, and there it remains, due to the existence of a surrounding large potential barrier. The potential well then starts to move, dragging the monomer attached to it. At any given time instant $t$, a single monomer is trapped by this potential and the whole process is completed when the time dependent potential gets back to its original state, now with a different monomer at the pore position. A single monomer translocation is accomplished at time $t=1/v$.

It is important to note that while dealing with expectation values such as
\be
\langle e^{-\beta W} \rangle\approx\frac{1}{N}\sum_{i=1}^{N}e^{-\beta W_{i}} \  \
\ee
in the JE, some special care is in order, since the large fluctuations displayed by $W$ enforces us to employ large statistical ensembles, which we have been able to produce thanks to a well-established cloud computing framework. Also, it is worth of remarking that the single monomer translocation event is a convenient choice to deal with, related to amounts of $W$ that are not large enough to ruin the precision of our statistical averages.

The work $W$ which is used in the JE is obtained by means of the following procedure:

(i) After the thermalization process, we turn off the stochastic and dissipation forces, so that only conservative forces play a role in the system and, consequently, the total energy variation is independent of the path. In this case, recalling that the external work done on the system is, by definition, its total mechanical energy variation, i.e., $W = \Delta E$, which takes into account all the interactions introduced in Sec. III and the time-dependent potential (\ref{Upot}) as well, we can assure that
\be
\langle e^{-\beta W}\rangle=\langle e^{-\beta \Delta E}\rangle=e^{-\beta \Delta F} \ . \
\label{freeenergy}
\ee

(ii) The time-dependent potential starts to act at $t=0$ ($\lambda =0$) and then stops at $t=1/v$ ($\lambda = \pi$) when the trapping potential (\ref{Upot}) recovers its original configuration and the monomer that follows the one that was previously attached to the pore takes the place of the latter.

We have produced polymer thermal equilibrium states by just following the behavior of its mechanical energy as a function of time in the Langevin simulations. More precisely, we evaluate the time scale $\tau_E$ for the polymer energy to reach a stationary statistical state, and, to make sure that the polymer is actually in thermal equilibrium, we work with polymer configurations taken only after the total evolution time $10 \tau_E$.

We have made our simulations to comply as close as possible with the assumptions put forward in the original discussion of the JE~\cite{Jarzynski}. That is why the thermal bath (modelled here through the stochastic and dissipative forces) is turned off during the single monomer translocation event. The great advantage in following the purest version of the JE is the complete suppression of errors associated to the thermal bath fluctuations.

Once we have established the procedure for the computation of free energy variations, we are now able to perform our simulations, which are run as follows. First, we fix the size $N$ of the polymer, say $N=50$, where initially half of the monomers are located in both sides of the membrane ($m=25$) in a situation of thermal equilibrium. After a period of time $t=1/v$, the translocation of a single monomer is deterministically driven towards the trans side of the membrane, so that the polymer ends up having $m=26$ monomers in that region. This procedure is repeated for $100$ independent realizations, in such a way that each one of these realizations takes the time interval $t=400$ ($v=2.5 \times 10^{-3}$) to be completed. From~(\ref{freeenergy}), we can obtain the variation of free energy for each individual translocation. This algorithm is then repeated for increasing values of $m$ to generate the free energy profiles related to the whole process of translocation.

\begin{figure}[center]
\includegraphics[scale=0.45]{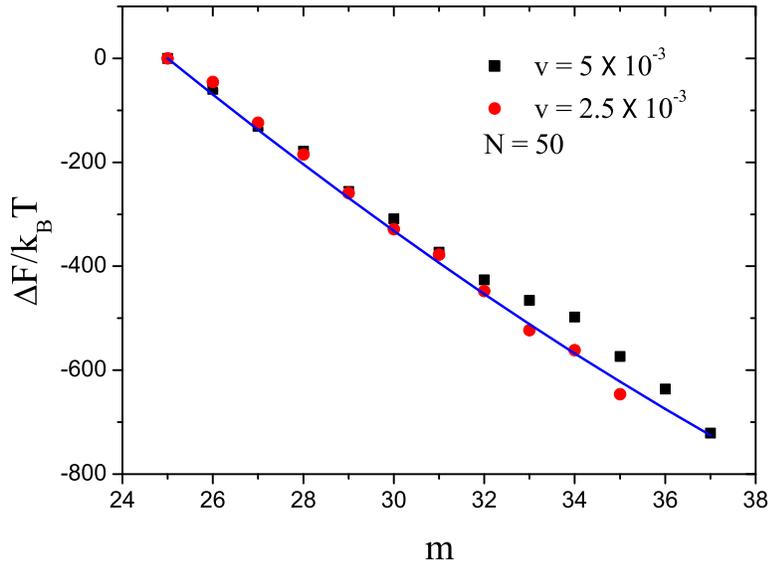}
\caption{The behavior of the free energy as a function of $m$, the number of monomers in the trans side of the membrane, as obtained through the JE applied to Langevin simulations, for velocities $v=5 \times 10^{-3}$ and $v=2.5 \times 10^{-3}$. In both cases, the size of the polymer was set to $N=50$. The solid line, which yields a reasonable interporlation of the data is obtained for a scaling exponent $\alpha=1.53$, as defined in Eq.~(\ref{aproxmuthu})}
\label{speed}
\end{figure}

Fig.~\ref{speed} illustrates the comparison, for a polymer of size $N=50$, between the cases of two different translocation velocities, $v=5 \times 10^{-3}$ and $v=2.5 \times 10^{-3}$. Our application of the JE is clearly validated, once we have found that both travel velocities of the potential lead to compatible free energy profiles; the case of smaller translocation velocity yields smaller fluctuations, as expected. We have found, as also shown in Fig.~\ref{speed}, that the free energy variations $\Delta F$ can be reasonably approximated as
\be
\frac{\Delta F}{k_{B}T}\approx A\left[(N-m)^{\alpha}-\left(\frac{N}{2}\right)^{\alpha}\right] \ , \
\label{aproxmuthu}
\ee
where $A$ and $\alpha$ are fitting constants, and $m$ is the number of monomers in the trans side of the membrane. $\Delta F$ is the free energy subtracted from its reference value defined at $m=N/2$. Good interpolations are obtained for the parameter values $A=1.0$ and $\alpha=1.53$.

\begin{figure}[center]
\includegraphics[scale=0.4]{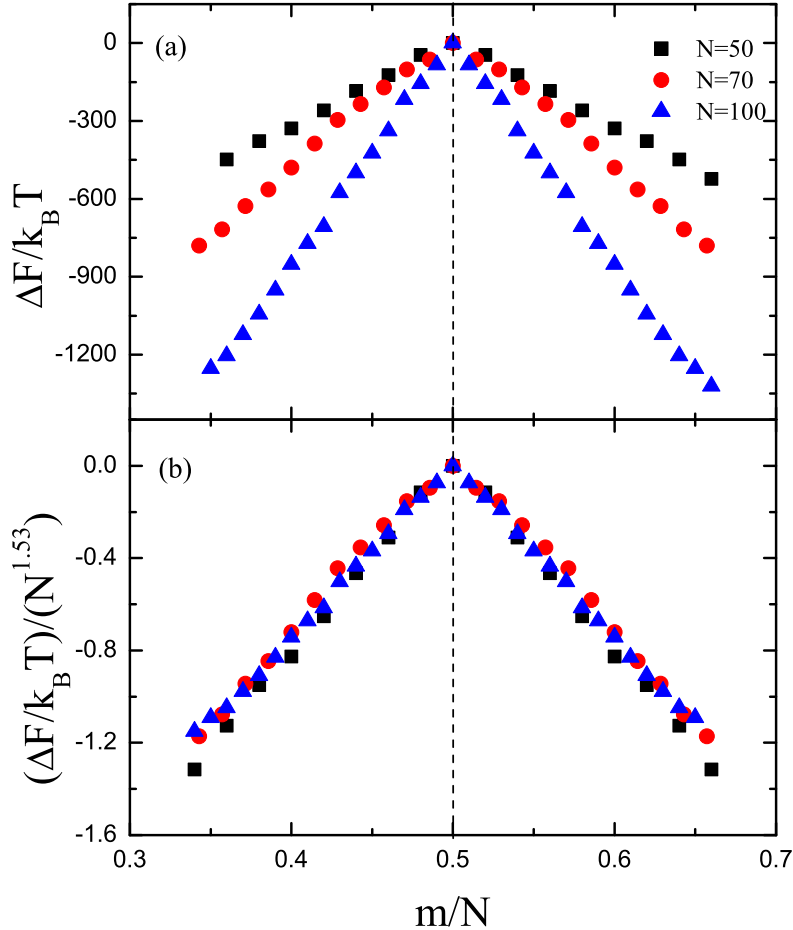}
\caption{In (a) we plot the free energy profiles obtained via JE as a function of the trans-side monomer concentration. We have worked with the polymer sizes $N=50$, $70$ and $100$. In (b) a clear finite-size scaling collapse is shown for the free energy values when they are rescaled by the size dependent factor $N^{1.53}$.}
\label{estimator}
\end{figure}

The JE-based procedure for obtaining free energy variations has been carried out for polymers of sizes $N=50$, $70$ and $100$, with the trapping potential at the pore moving with velocity $v=2.5 \times 10^{-3}$. The corresponding results are displayed in Fig.~\ref{estimator}, in which, furthermore, we check the finite size scaling behavior of the free energy indicated from Eq. (\ref{aproxmuthu}).

\begin{figure}[center]
\includegraphics[scale=0.4]{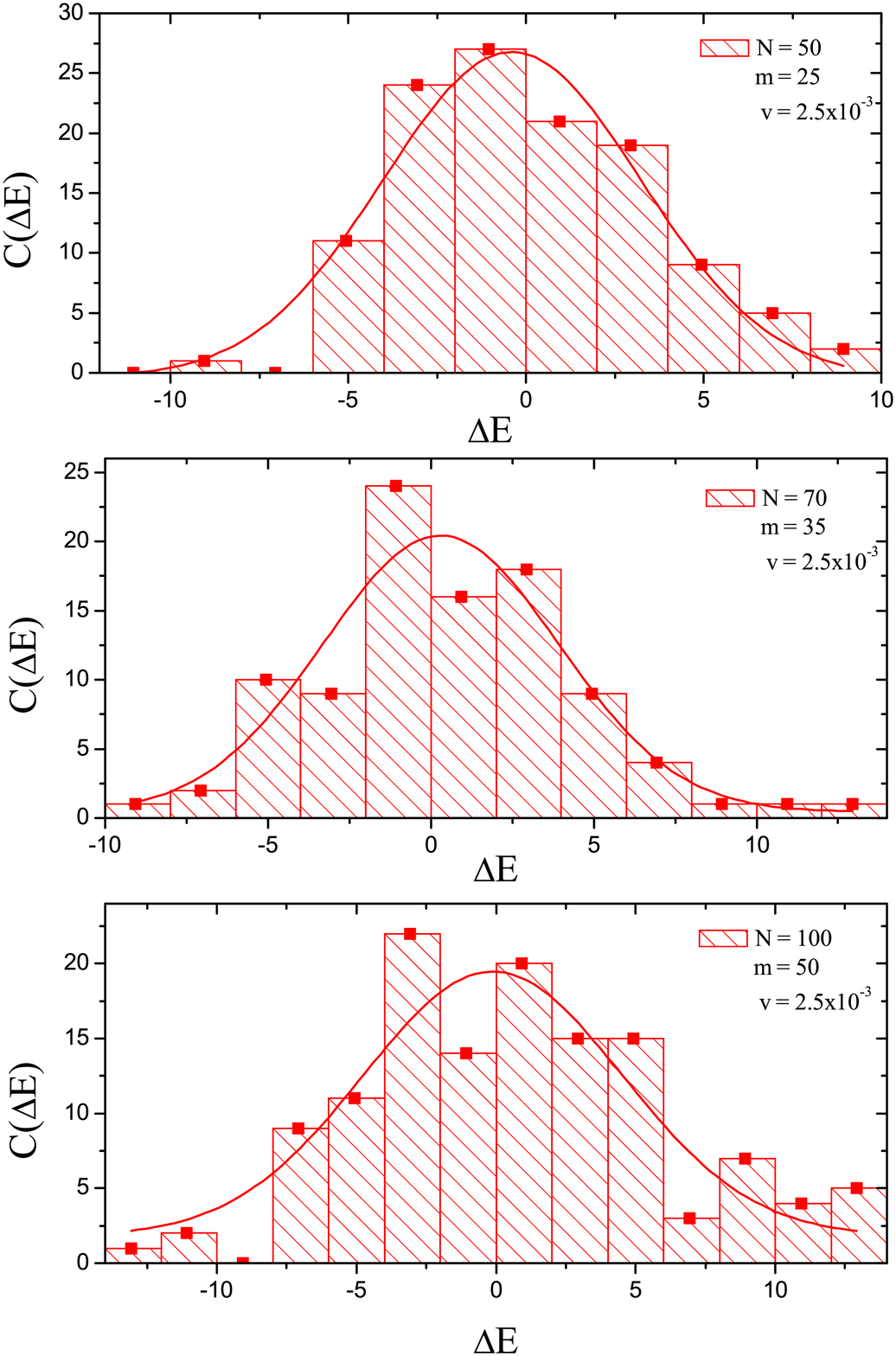}
\caption{Histograms for the work $W=\Delta E$ associated with single monomer translocation events. The solid lines are (non-normalized) gaussian distributions which have the same mean and variance as the ones computed from the histograms.}
\label{histograms}
\end{figure}

Illustrative work histograms for translocations with initial $m=N/2$ are given in Fig.~\ref{histograms}, for $N=50$, $70$, and $100$. As it can be clearly seen from Fig.~\ref{histograms}, we deal with out-of-equilibrium translocation events, which define the statistical ensembles used in the computation of the JE expectation values.

It is interesting to compare the results of Fig.~\ref{estimator} with the free energy profile described through Eq. (\ref{muthu2}). One clearly observes that the free energy (\ref{muthu2}) decays not so fastly as the ones we have computed do. The main point is that our model includes, {\it{ab-initio}}, the vibrating degrees of freedom, in contrast to the original formulation of Ref.~\cite{Muthukumar199910371}. The thermal effects associated to vibrations are likely, as suggested by our results, to strongly affect the free energy profile of the translocating polymer.

\begin{figure}[center]
\includegraphics[scale=0.4]{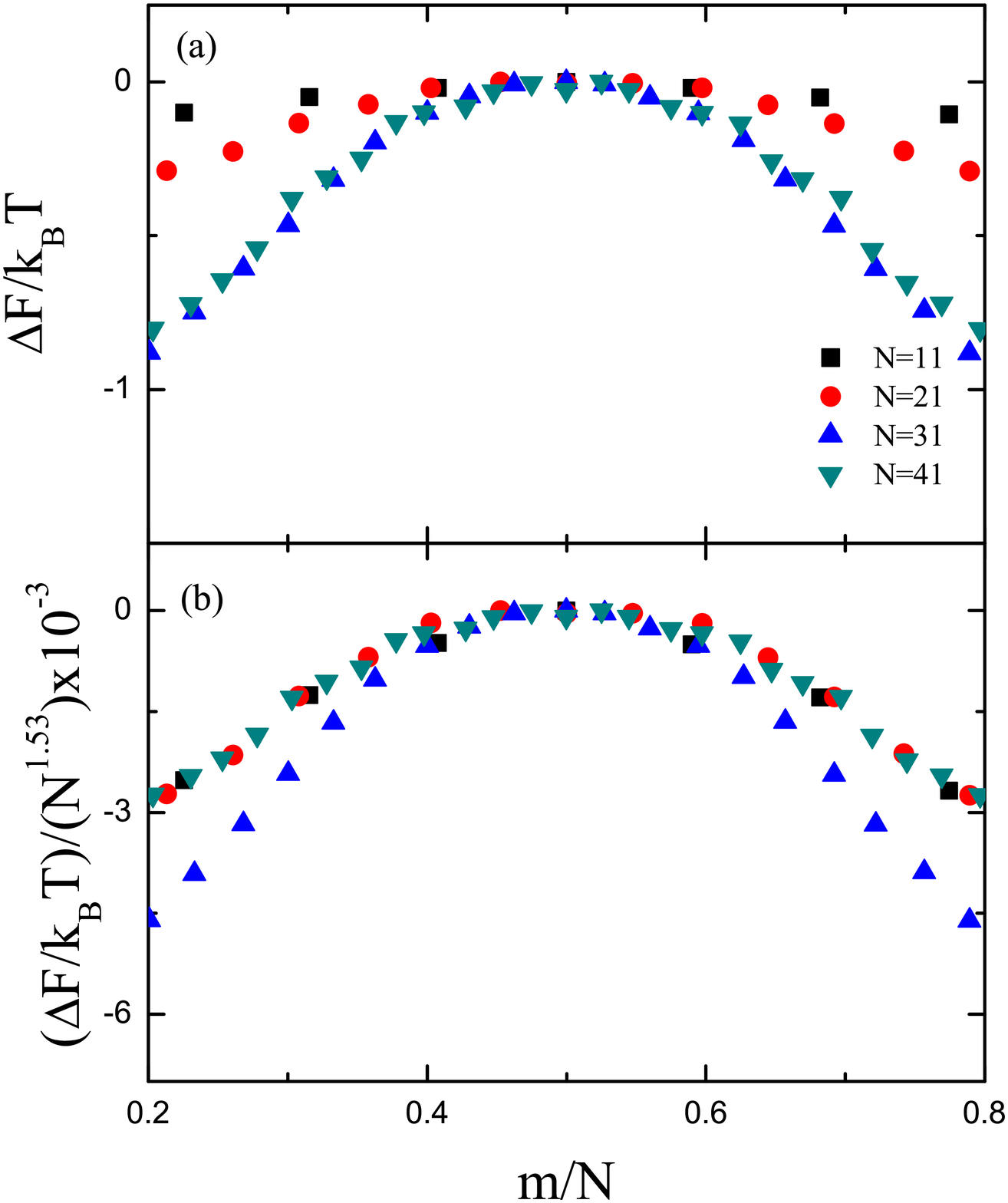}
\caption{(a) The numerical data reported in Fig. $4$(a) of Ref.~\cite{Li2012} is (b) alternatively
plotted taking into account a rescaling $N^{1.53}$ factor. The collapse is in fact similar to the one shown in our
Fig.~\ref{estimator}. We note that there is a clear anomaly of unknown origin for the case $N = 31$.}
\label{Lienergy}
\end{figure}

The collapse of the free energy profiles into a single curve, due to finite-size scaling effects, is a central result of our work. We call attention to a suggestive agreement between our results and the ones of Ref.~\cite{Li2012}, in which the free energy of a polymer interacting with a membrane is determined through histogram methods - a completely different methodological procedure - for relatively small polymer chains ($N \leq 41$). The results shown in~\cite{Li2012} indicate that as the chain length increases the free energy curve no longer displays the profile given in (\ref{muthu2}). Moreover, we have verified that the free energy data reported in Ref.~\cite{Li2012} have a finite- size scaling collapse similar to the one obtained here, as one can see from Fig.~\ref{Lienergy} (there seems to be, however, some anomaly with the data for polymers of size $N=31$).
To place our results into a proper context, we point out the need of further numerical work, devised for even larger polymer chains (at least up to $N=10^3$) and different temperatures, in order to check that the scaling behavior of the free energy, as addressed in Eq. (\ref{aproxmuthu}), is indeed asymptotically correct.

\vspace{0.2cm}
\section{Conclusion}

We have investigated, taking the JE as a fundamental tool, the free energy profiles for the problem of unbiased three-dimensional polymer translocation. The use of a large cloud computing resource has allowed us to work with polymers and statistical ensembles of sizes which go reasonably beyond the ones commonly found in the literature of the subject. In our Langevin simulations, the use of the JE has been implemented by means of a weakly invasive time dependent potential that behaves like an optical trap pulling monomers along a single monomer translocation event.

Our results show that the educated guess~(\ref{muthu2}) for the free energy profile~\cite{Muthukumar199910371} does not account with the behavior derived from Langevin simulations. As a clear result, we note that the larger is the size of the polymer the steeper the free energy becomes. Another important remark is the finite-size scaling collapse obtained for our data and for the one reported in Ref.~\cite{Li2012} for small polymer sizes, when the free energy is rescaled by the factor $N^{1.53}$. A challenging statistical mechanics problem, of course, is to understand the finite-size scaling properties of the free energy discussed in Sec. IV.

An interesting question, which is actually open to further study, is whether the Fokker-Planck approach~\cite{Muthukumar199910371} can lead to relevant results, if more realistic free energy profiles are taken into account. A motivating work, in this direction of research, is related to the discovery of free energy oscillations due to pore thickness effects~\cite{polson2013}. Fluctuations arising from the delay due to the interactions with the membrane atoms are not captured in our study because of the one-atom-thick membrane used in our simulations. Also, we have not considered in this work situations where the initial monomer position would be slightly displaced from the membrane pore, in order to provide a more detalied account of the free energy variations as monomers cross the membrane pore. As a matter of principle, all of the aforementioned problems can be numerically studied along lines which have been addressed in the present paper.

This work has been partially supported by CNPq and FAPERJ.

\bibliography{biblio}

\begin{thebibliography}{10}

\bibitem{suntharalingam2003}
M.~Suntharalingam and S.~R. Wente,
\newblock Developmental cell {\bf 4}, 775 (2003).

\bibitem{Tseng}
Y.-L. Tseng, J.-J. Liu, and R.-L. Hong,
\newblock Mol. Pharm. {\bf 62}, 864 (2002).

\bibitem{Szabo}
I.~Szabo et~al.,
\newblock J. Biol. Chem. {\bf 272}, 25275 (1997).

\bibitem{Hanss}
B.~Hanss, E.~Leal-Pinto, L.~A. Bruggeman, T.~D. Copeland, and P.~E. Klotman,
\newblock Proc. Natl. Acad. Sci. {\bf 95}, 1921 (1998).

\bibitem{Kasianowicz}
J.~Kasianowicz, E.~Brandin, D.~Branton, and D.~Deamer,
\newblock Proc. Natl Acad. Sci. USA {\bf 93}, 13770 (1996).

\bibitem{Grosberg}
A.~Grosberg, S.~Nechaev, M.~Tamm, and O.~Vasilyev,
\newblock Phys. Rev. Lett. {\bf 96} (2006).

\bibitem{Huopaniemi}
I.~Huopaniemi, K.~Luo, T.~Ala-Nissila, and Y.~S.-C.,
\newblock J. Chem. Phys. {\bf 125} (2006).

\bibitem{Luo2007}
K.~Luo, T.~Ala-Nissila, Y.~S.-C., and A.~Bhattacharya,
\newblock Phys. Rev. Lett. {\bf 99} (2007).

\bibitem{Wei2007}
D.~Wei, W.~Yang, X.~Jin, and Q.~Liao,
\newblock J. Chem. Phys. {\bf 126} (2007).

\bibitem{Luo2008}
K.~Luo, T.~Ala-Nissila, Y.~S.-C., and A.~Bhattacharya,
\newblock Phys. Rev. Lett. {\bf 100} (2008).

\bibitem{Bhattacharya2009}
A.~Bhattacharya,
\newblock Eur. Phys. J. E {\bf 29}, 423 (2009).

\bibitem{Gauthier2009}
M.~Gauthier and G.~Slater,
\newblock Phys. Rev. E {\bf 79} (2009).

\bibitem{Panja2007}
D.~Panja, G.~Barkema, and R.~Ball,
\newblock J. Phys. Condens. Matter {\bf 19} (2007).

\bibitem{Sung}
W.~Sung and P.~Park,
\newblock Phys. Rev. Lett. {\bf 77}, 783 (1996).

\bibitem{Muthukumar199910371}
M.~Muthukumar,
\newblock J. Chem. Phys. {\bf 111}, 10371 (1999).

\bibitem{Chuang2002}
J.~Chuang, Y.~Kantor, and M.~Kardar,
\newblock Phys. Rev. E {\bf 65}, 011802/1 (2002).

\bibitem{Mondaini2}
F.~Mondaini and L.~Moriconi,
\newblock Phys. Lett. A {\bf 376}, 2903 (2012).

\bibitem{Dubbeldam2007}
J.~Dubbeldam, A.~Milchev, V.~Rostiashvili, and T.~Vilgis,
\newblock Phys. Rev. E {\bf 76} (2007).

\bibitem{sakaue2007}
T.~Sakaue,
\newblock Phys. Rev. E {\bf 76}, 021803 (2007).

\bibitem{sakaue2010}
T.~Sakaue,
\newblock Phys. Rev. E {\bf 81}, 041808 (2010).

\bibitem{rowghanian2011}
P.~Rowghanian and A.~Y. Grosberg,
\newblock J. Phys. Chem. B {\bf 115}, 14127 (2011).

\bibitem{lehtola2009}
V.~Lehtola, R.~Linna, and K.~Kaski,
\newblock EPL {\bf 85}, 58006 (2009).

\bibitem{lehtola2008}
V.~Lehtola, R.~Linna, and K.~Kaski,
\newblock Phys. Rev. E {\bf 78}, 061803 (2008).

\bibitem{polson2013}
J.~M. Polson, M.~F. Hassanabad, and A.~McCaffrey,
\newblock J. Chem. Phys. {\bf 138}, 024906 (2013).

\bibitem{Jarzynski}
C.~Jarzynski,
\newblock Phys. Rev. Lett. {\bf 78}, 2690 (1997).

\bibitem{Latinwo}
F.~Latinwo and C.~M. Schroeder,
\newblock Soft Matter {\bf 10}, 2178 (2014).

\bibitem{Seifert}
U.~Seifert,
\newblock Rep. Prog. Phys. {\bf 75}, 126001 (2012).

\bibitem{teste}
Our statistical data sets have been produced within the Grid Initiatives for
  e-Science virtual communities in Europe and Latin America (GISELA), a cloud
  computing framework supported by several academic institutions which
  comprises up to 10.000 processors.

\bibitem{Bena2005}
I.~Bena, C.~V. den Broeck, and R.~Kawai,
\newblock EPL {\bf 71}, 879 (2005).

\bibitem{Blicke2006}
V.~Blickle, T.~Speck, L.~Helden, U.~Seifert, and C.~Bechinger,
\newblock Phys. Rev. Lett. {\bf 96}, 070603 (2006).

\bibitem{Liphardt2002}
J.~Liphardt, S.~Dumont, S.~B. Smith, I.~Tinoco, and C.~Bustamante,
\newblock Science {\bf 296}, 1832 (2002).

\bibitem{Hummer2001}
G.~Hummer and A.~Szabo,
\newblock Proc. Natl. Acad. Sci. {\bf 98}, 3658 (2001).

\bibitem{lennard1937}
J.~Lennard-Jones and A.~Devonshire,
\newblock Proc. R. Soc. A {\bf 163}, 53 (1937).

\bibitem{Li2012}
X.~Xiang-Gui, Z.~Li, L.~Zhong-Yuan, and L.~Ze-Sheng,
\newblock Phys. Lett. A {\bf 376}, 290 (2012).

\end{thebibliography}
\bibliographystyle{phaip}
\end{document}